\documentclass[twocolumn, 10pt]{IEEEtran}

\usepackage{hyperref,bm,graphicx,amsmath,amssymb}
\usepackage{algorithm,algpseudocode}

\usepackage{color, soul}

\begin{document}
\title{ Phase Transition Analysis of Sparse Support Detection from Noisy Measurements }
\setlength{\baselineskip}{1.0\baselineskip}

\author{ Jaewook~Kang,
        Heung-No~Lee,
        and~Kiseon~Kim\\

\bigskip
\IEEEauthorblockA{School of Information and Communication,\\
 Department of Nanobio Materials and Electronics,\\
 Gwangju Institute of Science and Technology (GIST), Republic of Korea} \vspace{-20pt}

\thanks{This work was supported by the World-Class University Program
(R31-10026), Haek-Sim Research Program (NO. 2012-047744), Do-Yak
Research Program (NO.2012-0005656), and Leading Foreign Research
Institute Recruitment Program (K20903001804-11E0100-00910) through
the National Research Foundation of Korea funded by the Ministry of
Education, Science, and Technology (MEST)}
 }\maketitle
\vspace{-20pt}

\begin{abstract}
This paper investigates the problem of sparse support detection
(SSD) via a detection-oriented algorithm named Bayesian hypothesis
test via belief propagation (BHT-BP)  \cite{SSP2012},\cite{BHT-BP}.
Our main focus is to compare BHT-BP to an estimation-based
algorithm, called CS-BP \cite{CS-BP}, and show its superiority in
the SSD problem. For this investigation, we perform a phase
transition (PT) analysis over the plain of the noise level and
signal magnitude on the signal support. This PT analysis sharply
specifies the required signal magnitude for the detection under a
certain noise level.  In addition, we provide an experimental
validation to assure the PT analysis. Our analytical and
experimental results show the fact that BHT-BP detects the signal
support against additive noise more robustly than CS-BP does.

\end{abstract}

\begin{keywords}
Noisy sparse recovery, sparse support detection, phase transition,
belief propagation.
\end{keywords}

\section{Introduction}
Noisy sparse recovery is referred to the problem  of recovering a
sparse signal $\mathbf{x}_0 \in \mathbb{R}^N$ from noisy linear
projection $\mathbf{y} \in \mathbb{R}^M$ generated by an
underdetermined system $(M<N)$.  In such a problem, sparse support
detection (SSD) is important because once the support is known, the
uncertainty to the recovery is confined to the additive noise, which
can be optimally solved in terms of mean squared errors using the
simple least square approach \cite{tropp}. Nevertheless, most
recovery algorithms to date for the problem have been developed
under the auspices of signal estimation rather than support
detection. They include algorithms developed under the criteria of
the $l_1$-norm minimization and the MAP-estimation such as
\emph{least absolute shrinkage and selection operator} (LASSO)
\cite{Lasso} and \emph{Baysian compressive sensing via belief
propagation} (CS-BP)  \cite{CS-BP}, respectively.

Thus, we make note of the fact that these estimation-based
algorithms may not be good choices when it comes to the SSD problem
under noisy setup. Indeed, recently several studies have indicated
that the existing estimation-based algorithms lead to a potentially
large gap with respect to the theoretical limit for the noisy
support recovery \cite{Wainwright2}-\cite{Fletcher}. Wainwright and
Akcakaya \emph{et al.} have shown that LASSO, a sparse signal
estimator, has a significant performance gap from the Fano's
detector which is  information-theoretically optimal, in the linear
sparsity regime \cite{Wainwright2},\cite{Akcakaya}. Furthermore,
Fletcher \emph{et al.} noted the suboptimality of LASSO and OMP
compared to the maximum likelihood detector and they remarked that
the gap grows as SNR increases \cite{Fletcher}.

In this paper, we consider a recently proposed detection-oriented
algorithm named \emph{Bayesian hypothesis test via belief
propagation} (BHT-BP) \cite{SSP2012},\cite{BHT-BP}. This algorithm
detects the signal support through a sequence of binary hypothesis
tests, where each hypothesis test is designed from the MAP-detection
criterion using Bayesian philosophy. In our previous studies
\cite{SSP2012},\cite{BHT-BP}, we have shown in extensive simulation
that BHT-BP works better than the estimation-based algorithms, such
as CS-BP and LASSO, for the SSD problem, particularly when
situations are noisy. In addition, we noted that BHT-BP is
noteworthy as a low-computational algorithm having $O(N \log N)$
order of complexity, enabled by belief propagation (BP) working on
sparse measurement matrices. In those studies, however, the
superiority of BHT-BP to the estimation-based algorithms was
verified only in simulation.


The main focus of this paper is to introduce a phase transition (PT)
analysis which can be used to describe how the support detection of
the BHT-BP algorithm behaves as the additive noise level is varied.
Namely, it provides an exact border line between success and failure
of the algorithm on the plane of the noise level and the signal
magnitude. The term ``phase transition" was first used by Donoho and
Tanner \cite{PT1},\cite{PT2} in the sparse recovery literature,
where   the relation between the undersampling ratio and the signal
sparsity was the focus which is different from the work studied
here.

The importance of the signal magnitude, the smallest magnitude on
the support to be more precise, in noisy sparse recovery problems
has been emphasized in  \cite{Wainwright2},\cite{Fletcher} where
they have shown that the required number of compressive measurements
for support detection is inversely proportional to the power of the
smallest magnitude on the support. Their study, however, did not
answer the following question: for successful support detections,
how large should the signal magnitude be at a fixed noise level?
With the PT analysis in this paper, such a statement can be made
sharp.  In addition, in order to verify the superiority of the
detection-oriented algorithm BHT-BP to the estimation-based one
CS-BP, we compare the PT region of BHT-BP to that of CS-BP. The
comparisons are made both in analysis and simulations which confirm
that the PT region of BHT-BP is larger than that of CS-BP.

\begin{figure*}[!t]
\centering
\includegraphics[width=12cm]{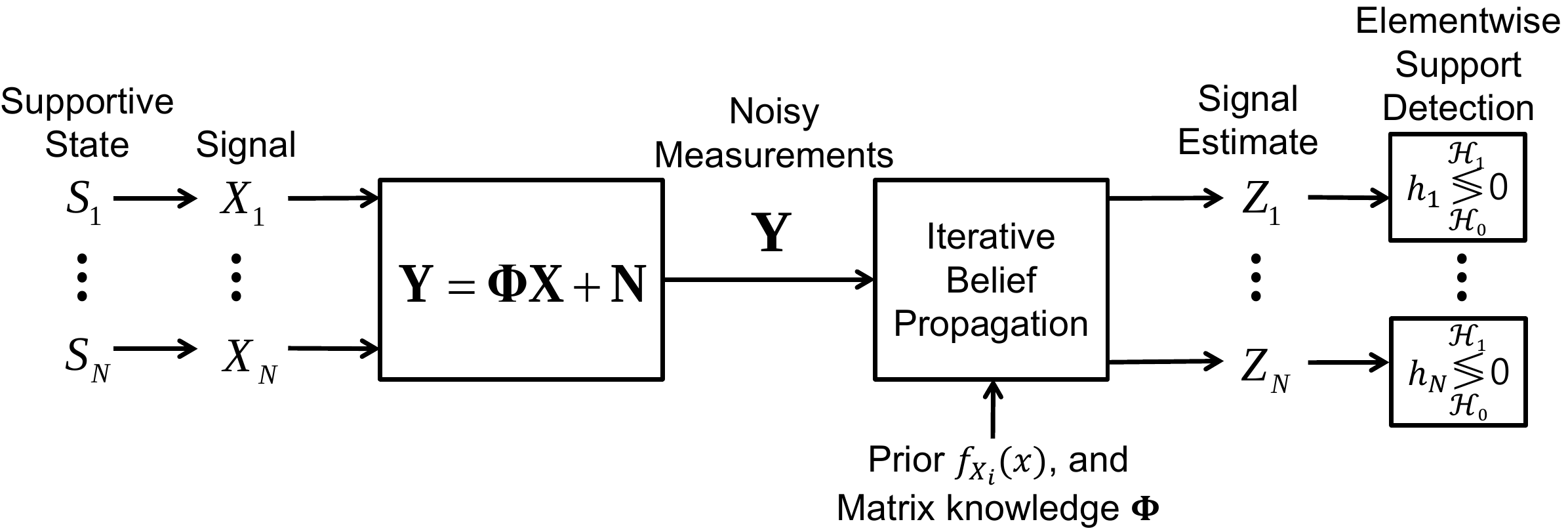}
\caption{Channel model for  sparse support detection}
\label{fig:fig2-1}
\end{figure*}

\section{Problem Formulation}
\subsection{Signal Model}
We consider a sparse signal $\mathbf{x}_0 \in \mathbb{R}^N$ which is
a  realization of a random vector $\mathbf{X}$. Here we assume that
the elements of $\mathbf{X}$ are \emph{i.i.d.} and that the
supportive state of each $X_i$ is determined by a Bernoulli random
variable $S_i$ with  a mixing rate $q:=\Pr\{S_i = 1\}$, \emph{i.e.},
\begin{eqnarray}\label{eq:eq2-1}
{S_i} = \left\{ \begin{array}{l}
1,\,\,\,\,\,{\text{if}}\,\,\ X_i \ne 0\\
0,\,\,\,\,\,{\text{else}}
\end{array} \right.\forall\,\, i \in \{1,...,N\}.
\end{eqnarray}
Hence, the state vector $\mathbf{S} \in \{0,1\}^N$ fully contains
the support information on the signal $\mathbf{X}$. For nonzero
values, we consider equiprobable signed values \emph{i.e.},
$X_{i|S_i=1} \sim \frac{1}{2}{\delta _{ + {x}}} + \frac{1}{2}{\delta
_{ - {x}}}$ where $\delta_{\tau}$ denotes the delta function peaked
at $\tau$.

Then, the support detector observes a noisy measurement vector,
\emph{i.e.},
\begin{align}\label{eq:eq2-2}
\mathbf{y}=\mathbf{\Phi}\mathbf{x}_0+\mathbf{w} \in \mathbb{R}^M,
\end{align}
where we  use  sparse measurement matrices $\mathbf{\Phi} \in
\{0,1\}^{M \times N}$ whose matrix sparsity is regulated by its
fixed column weight, \emph{i.e.}, $ \left| \{ j: \phi_{ji} \ne 0\}
\right|=L$, and a zero-mean Gaussian vector $\mathbf{w}\in
\mathbb{R}^M$ drawn from $\mathcal{N}(0,\sigma_W^2\mathbf{I})$ to
represent additive white noise.

In the SSD problem, we assume a spike-and-slab prior such that for
each signal element $X_i$, it is given by
\begin{eqnarray} \label{eq:eq2-3}
\begin{array}{l}
f_{X_i}(x):= qf_{X_i}( x |S_i = 1)+ (1 -
q)f_{X_i}( x |S_i = 0)\\
\,\,\,\,\,\,\,\,\,\,\,\,\,\,\,\,\,\,\,= q\mathcal{N}(x;0,\sigma
_{X}^2 )+ (1 - q)\delta_0,
\end{array}
\end{eqnarray}
where we use $f(\cdot)$ to denote a probability density function
(PDF). The variance $\sigma_X^2$ is calibrated according the
magnitude of the target signal $\mathbf{x}_0$. Namely, if the signal
has large values on its support, the variance $\sigma_X^2$ should be
sufficiently large for the support detection.

\subsection{Channel Model}
Signal support detection  can be performed in an elementwise manner
on the basis of \emph{decoupling principle}
\cite{Guo05},\cite{Guo07}. According to this principle, the vector
measurement channel can be decoupled to a sequence of scalar
Gaussian channels in the large system limit ($N \rightarrow
\infty$). In this case, the input of each elementwise support
detection is a scalar estimate of $X_i$ denoted by $Z_i$, as shown
in Fig.\ref{fig:fig2-1}.

The decoupling for the elementwise detection can be achieved using
an iterative BP algorithm. Indeed, Guo and Wang have shown that the
BP algorithm  finds the marginal posteriors of the signal exactly if
the matrix $\mathbf{\Phi}$ is assumed to be a sparse matrix with the
\emph{no-short-cycle} property \cite{Guo06}. Such a setting is
called  \emph{large-sparse-system} (LSS) setup. Fig.\ref{fig:fig2-1}
shows the overall channel model considered in this paper in which we
assume that the signal length $N$ is sufficiently large to justify
the LSS setup.

\subsection{Problem Statement}
In this study, we aim to analytically compare the  noisy behavior of
the two types of algorithms:  BHT-BP  and CS-BP. The CS-BP algorithm
is an estimation-based algorithm which obtains the signal estimate
directly from the posterior density of $\mathbf{Z} \in \mathbb{R}^N$
using the MAP or the MMSE estimator \cite{CS-BP}. Therefore, in
CS-BP, the support information is provided as a by-product of the
signal estimate. Namely, if a scalar estimate given by $Z_i$ is
nonzero, then the algorithm simply decides the element belongs to
the signal support. Accordingly, the detection function for a
supportive state $S_i$ in CS-BP is described by
\begin{align}\label{eq:eq2-7}
h_{\text{CS-BP},i}:=\log \frac{{{f_{{X_i}}}(x = {{\widehat
x}_{\text{MAP},i}}|{Z_i},\mathbf{\Phi})}}{{{f_{{X_i}}}(x =
0|{Z_i},\mathbf{\Phi})}} \mathop {\mathop \gtrless
\limits_{{\mathcal{H}_0}} }\limits^{{\mathcal{H}_1}} 0,
\end{align}
where  $\mathcal{H}_0:=\{S_i=0\}$ and $\mathcal{H}_1:=\{S_i=1\}$
denote two possible hypotheses, and we assume CS-BP uses the
MAP-estimator such that an estimate of $X_i$ is given as ${\widehat
x_{\text{MAP},i}}: = \arg \mathop {\max }\limits_{x \ne 0}
{f_{{X_i}}}(x|Z_j,{\bf{\Phi }})$. The detection in \eqref{eq:eq2-7}
can be achieved with the marginal posterior
${f_{{X_i}}}(x|\mathbf{Y},{\bf{\Phi }})$ since
\begin{align}\label{eq:eq2-8}
{f_{{X_i}}}(x|{Z_j},{\bf{\Phi }}) = \int\limits_{\bf{Y}} {
\underbrace {{f_{{X_i}}}(x|{Z_j},{\bf{Y}},{\bf{\Phi }})}_{ =
{f_{{X_i}}}(x|{\bf{Y}},{\bf{\Phi
}})}{f_{\bf{Y}}}({\bf{y}}|{Z_i},{\bf{\Phi }})d{\bf{y}}},
\end{align}
and $f_{\bf{Y}}({\bf{y}}|{Z_i},{\bf{\Phi }}) \geq 0$ for all
$\mathbf{y}$ where $f_{X_i}(x|\mathbf{Y},Z_i,{\bf{\Phi
}})=f_{X_i}(x|\mathbf{Y},{\bf{\Phi }})$ is true because $Z_i$ is a
function of $\mathbf{Y}$. Therefore, \eqref{eq:eq2-7} can be
rewritten as
\begin{align}\label{eq:eq2-9}
h_{\text{CS-BP},i}=\log \frac{{{f_{{X_i}}}(x = {{\widehat
x}_{\text{MAP},i}}|{\bf{Y}},\mathbf{\Phi})}}{{{f_{{X_i}}}(x =
0|{\bf{Y}},\mathbf{\Phi})}} \mathop {\mathop \gtrless
\limits_{{\mathcal{H}_0}} }\limits^{{\mathcal{H}_1}} 0.
\end{align}
In contrast, for a detection-oriented algorithm BHT-BP, finding the
sparse support set is an end in itself. Therefore, the  detection
function of BHT-BP is designed from the MAP-detection of the
supportive state $S_i$, given by
\begin{align}\label{eq:eq2-4}
h_{\text{MAP},i}:=\log \frac{{\Pr \{ {S_i} = 1| Z_i,{\bf{\Phi }} \}
}}{{\Pr \{ {S_i} = 0|Z_i,{\bf{\Phi }}\} }} \mathop {\mathop \gtrless
\limits_{{\mathcal{H}_0}} }\limits^{{\mathcal{H}_1}} 0.
\end{align}
In \eqref{eq:eq2-4}, the posterior probability of a supportive state
$S_i$ can be decomposed as
\begin{align}\label{eq:eq2-5}
&\Pr \{ {S_i}|{Z_i},{\bf{\Phi }}\}  \nonumber\\
&= \int\limits_{{X_i},{\bf{Y}}}  \underbrace{\Pr \{
{S_i}|{X_i},{\bf{Y}},{Z_i},{\bf{\Phi }}\}}_{=\Pr \{
{S_i}|{X_i},{\bf{\Phi }}\}}  \underbrace
{{f_{{X_i}}}(x|{Z_j},{\bf{Y}},{\bf{\Phi }})}_{ =
{f_{{X_i}}}(x|{\bf{Y}},{\bf{\Phi
}})}\times\nonumber \\
& \,\,\,\,\,\,\,\,\,\,\,\,\,\,\,\,\,\times   {f_{\bf{Y}}}({\bf{y}}|{Z_i},{\bf{\Phi }})dxd{\bf{y}}\nonumber \\
&= \int\limits_{\bf{Y}} {\left[ {\int\limits_{{X_i}}
{\frac{{{f_{{X_i}}}(x|{S_i})\Pr \{ {S_i}\}
{f_{{X_i}}}(x|{\bf{Y}},{\bf{\Phi }}) }}{{{f_{{X_i}}}(x)}} dx} }
\right]{f_{\bf{Y}}}({\bf{y}}|{Z_i},{\bf{\Phi }})}
 d{\bf{y}},
\end{align}
where $\Pr \{ {S_i}|{X_i},{\bf{Y}},{Z_i},{\bf{\Phi }}\}=\Pr \{
{S_i}|{X_i},{\bf{\Phi }}\}$ holds true because $Z_i$ and
$\mathbf{Y}$  are conditionally independent of $S_i$ given $X_i$. In
addition, we know from \eqref{eq:eq2-8}  that
$f_{X_i}(x|\mathbf{Y},Z_i,{\bf{\Phi
}})=f_{X_i}(x|\mathbf{Y},{\bf{\Phi }})$. Therefore, the
MAP-detection can be achieved by considering only the integral
within brackets since $f_{\bf{Y}}({\bf{y}}|{Z_i},{\bf{\Phi }}) \geq
0$ for all $\mathbf{y}$. Using these facts, the detection function
of BHT-BP is defined as
\begin{align}\label{eq:eq2-6}
&h_{\text{BHT-BP},i}:=\log \frac{q}{1-q} + \log {\frac{{\int {
\frac{{f_{X_i}( x |S = 1) }}{{f_{X_i}( x ) }}
 f_{X_i}(x | {\bf{Y}},{\bf{\Phi }} )}dx }}{ {\int { \frac{{f_{X_i}( x |S = 0) }}{{f_{X_i}( x ) }} f_{X_i}(x | {\bf{Y}},{\bf{\Phi }} ) }dx } }} \mathop
{\mathop \gtrless \limits_{{\mathcal{H}_0}}
}\limits^{{\mathcal{H}_1}} 0.
\end{align}
These detection functions in \eqref{eq:eq2-9} and \eqref{eq:eq2-6}
will be used later on for development of the PT analysis in Section
III-B.

 As the first step of this analysis, we provide an analytical
expression of the marginal posterior
${f_{{X_i}}}(x|\mathbf{Y},{\bf{\Phi }})$ obtained via BP, under the
LSS setup. This posterior expression is used to represent the
detection function of CS-BP \eqref{eq:eq2-9} and  BHT-BP
\eqref{eq:eq2-6} as a function of $\sigma_W$ and $x_{0,i}$. Then,
using this result, we analyze the  failure event  for each detection
function and compare the condition for these two events over the
plane of the noise level $\sigma_W$ and signal magnitude $x_{0,i}$.
Our analytical and experimental results show that BHT-BP  detects
the signal support more robustly than CS-BP, for the given noise
level $\sigma_W$ and signal magnitude $|x_{0,i}|$.

\section{Main Result}
\subsection{Derivation of Marginal Signal Posterior}
Let ${V^{(l)}_{{X_i} \to {Y_j}}}$ and ${U^{(l)}_{{Y_j} \to {X_i}}}$
represent the BP-messages passed from $X_i$ to $Y_j$ and from $Y_j$
to $X_i$, respectively, for all pairs of $(i,j):\phi_{ji} \neq 0$ at
the $l$-th iteration. Our derivation starts from  the message
${U^{(l)}_{{Y_j} \to {X_i}}}$, expressed as
\begin{align}
{U^{(l)}_{{Y_j} \to {X_i}}}=Y_j-
 \sum\limits_{\scriptstyle k:{\phi _{jk}} \ne 0,k \ne i}{{\bf{E}}\{ {X_k}|{V^{(l-1)}_{{X_k} \to {Y_j}}},{\bf{\Phi }}\} }
+N_j.
\end{align}
By the central limit theorem (CLT), as a sum of i.i.d. random
variables, the message ${U_{{Y_j} \to {X_i}}}$   is asymptotically
Gaussian under the LSS setup \cite{Guo06}. Then, we have
\begin{align} \label{eq:eq3-1}
{U^{(l)}_{{Y_j} \to {X_i}}} \mathop \sim \limits^{{\text{by CLT}}}
\mathcal{N}\left(y_j-
 \sum\limits_{k}{{\bf{E}}\{ {X_k}|{V^{(l-1)}_{{X_k} \to {Y_j}}},{\bf{\Phi }}\}
 },\sigma_W^2+\sigma_{\text{CEI}}^2\right),
\end{align}
where $\sigma_{\text{CEI}}^2$ denotes  the variance of the
cross-element-interference (CEI). In addition, we use the fact that
the mean of the Gaussian PDF in \eqref{eq:eq3-1} converges to the
true value, \emph{i.e.}, $y_j- \sum\limits_{k}{{\bf{E}}\{
{X_k}|{V^{(l-1)}_{{X_k} \to {Y_j}}},{\bf{\Phi }}\} } \rightarrow
x_{0,i}$, and the interference term will be eliminated, \emph{i.e.},
$\sigma_{\text{CEI}}^2 \rightarrow 0$, as the iteration becomes
deeper $l \rightarrow \infty$ under the large system limit ($N
\rightarrow \infty$) \cite{Guo06}. For the convergence of BP, we
assume that the  mixing rate $q$ is very small ($0<q \ll 1$) such
that the signal is sparse enough. Accordingly, the PDF of messages
from the measurement side toward $X_i$ converges to a Gaussian PDF
with the mean $x_{0,i}$ and the variance $\sigma_W^2$, \emph{i.e.},
\begin{align}\label{eq:eq3-2}
f_{{U_{{Y_j} \to {X_i}}}}(u|X_{i},\mathbf{Y},{\bf{\Phi
}})\rightarrow \mathcal{N}(u;x_{0,i},\sigma_W^2).
\end{align}
As we discussed in Section II-B, the \emph{no-short-cycle} property
of the matrix $\mathbf{\Phi}$ ensures that elements of $\{ {U_{{Y_j}
\to {X_i}}}\}_{(i,j):\phi_{ji}\ne 0}$ are i.i.d. Then, using this
property and the Bayesian rule, the marginal posterior of each $X_i$
is obtained as
\begin{align} \label{eq:eq3-3}
f_{X_{i}}(x|\mathbf{Y},{\bf{\Phi }})& \mathop  \equiv \limits^{{\text{by BP}}} {f_{{X_i}}}(x|{\bf{Y}},{\bf{\Phi }},{\bf{U}}) \\
& = \eta \left[ f_{X_i}(x) \times \prod\limits_{j: \phi_{ji} \ne 0}
f_{ {U_{{Y_j} \rightarrow {X_i}}} } (u|X_i,\mathbf{Y},{\bf{\Phi }})
\right]\nonumber,
\end{align}
where   $\eta[\cdot]$ is the normalization function ensuring $\int
{f_{X_{i}}(x|\mathbf{Y},\mathbf{\Phi})dx =1}$. In addition, note
that $| \{ j: \phi_{ji} \ne 0\} |=L$ from our signal
model; hence, the product in \eqref{eq:eq3-3} is made up of $L$
densities for ${U_{{Y_j} \rightarrow {X_i}}} \,\,\forall j:
\phi_{ji} \neq 0$. By applying the prior knowledge given in
\eqref{eq:eq2-3} and the result of \eqref{eq:eq3-2} to
\eqref{eq:eq3-3}, we have
\begin{align}\label{eq:eq3-4}
&f_{X_{i}}(x|\mathbf{Y},{\bf{\Phi }})\nonumber\\
&=\eta \left[ q c_2 \mathcal{N}(x;
\frac{Lx_{0,i}\sigma_{X}^2}{L\sigma_{X}^2+\sigma_W^2},
\frac{\sigma_{X}^2\sigma_W^2}{L\sigma_{X}^2+\sigma_W^2}) +
(1-q)c_1\delta_0 \right],
\end{align}
where we use the fact that the product of Gaussian PDFs results in a
scaled Gaussian PDF, \emph{i.e.},
\begin{eqnarray}\label{eq:eq3-5}
\begin{array}{l}
\mathcal{N}(x;{\mu _1},\sigma _1^2) \times \mathcal{N}(x;{\mu
_2},\sigma _2^2) \propto \mathcal{N}(x;b,B),\nonumber
\end{array}
\end{eqnarray}
with $B = \frac{{\sigma _1^2\sigma _2^2}}{{\sigma _1^2 + \sigma
_2^2}},b = \frac{{{\mu _1}\sigma _2^2 + {\mu _2}\sigma
_1^2}}{{\sigma _1^2 + \sigma _2^2}}$. Hence, the constants $c_1,c_2$
in \eqref{eq:eq3-4} are defined by
\begin{equation*}
\begin{array}{l}
c_1:= \exp\left[-\frac{Lx_{0,i}^2}{2\sigma_W^2} \right]/
\sqrt{2\pi\sigma_W^2/L},
\end{array}
\end{equation*}
and
\begin{equation*}
\begin{array}{l}
{c_2}:= {{\exp \left[ {\frac{{x_{0,i}^2 }}{{2(\sigma_{X}^2 +
\sigma_W^2/L)} }} \right]}}/{{\sqrt {2\pi (\sigma_{X}^2 +
\sigma_W^2/L)} }}.
\end{array}
\end{equation*}
The expression in \eqref{eq:eq3-4} reveals that the marginal
posterior  consists of a slab-PDF $\mathcal{N}(x;
\frac{Lx_{0,i}\sigma_{X}^2}{L\sigma_{X}^2+\sigma_W^2},
\frac{\sigma_{X}^2\sigma_N^2}{L\sigma_{X}^2+\sigma_W^2})$ and a
zero-spike $\delta_0$. For  convenience, we define the mixing rate
$\rho_i$, the mean $\mu_i$, and the variance $\theta_i^2$  of the
marginal posterior as $\rho_{i}:=\frac{q c_2}{q c_2+ (1-q) c_1}$,
$\mu_{i}:=\frac{Lx_{0,i}\sigma_{X}^2}{L\sigma_{X}^2+\sigma_W^2}$,
and
$\theta_{i}^2:=\frac{\sigma_{X}^2\sigma_N^2}{L\sigma_{X}^2+\sigma_W^2}$,
respectively. Then, using these parameters $\rho_i, \mu_i,
\theta_i^2$, we can rewrite the marginal posterior in
\eqref{eq:eq3-4} as a spike-and-slab PDF, \emph{i.e.},
\begin{eqnarray}\label{eq:eq3-7}
f_{X_{i}}(x|\mathbf{Y},{\bf{\Phi }})=\rho_{i} \mathcal{N}(x;
\mu_{i}, \theta_{i}^2)+ (1-\rho_{i})\delta_0.
\end{eqnarray}

\begin{figure*}[!t]
\centering
\includegraphics[width=9cm]{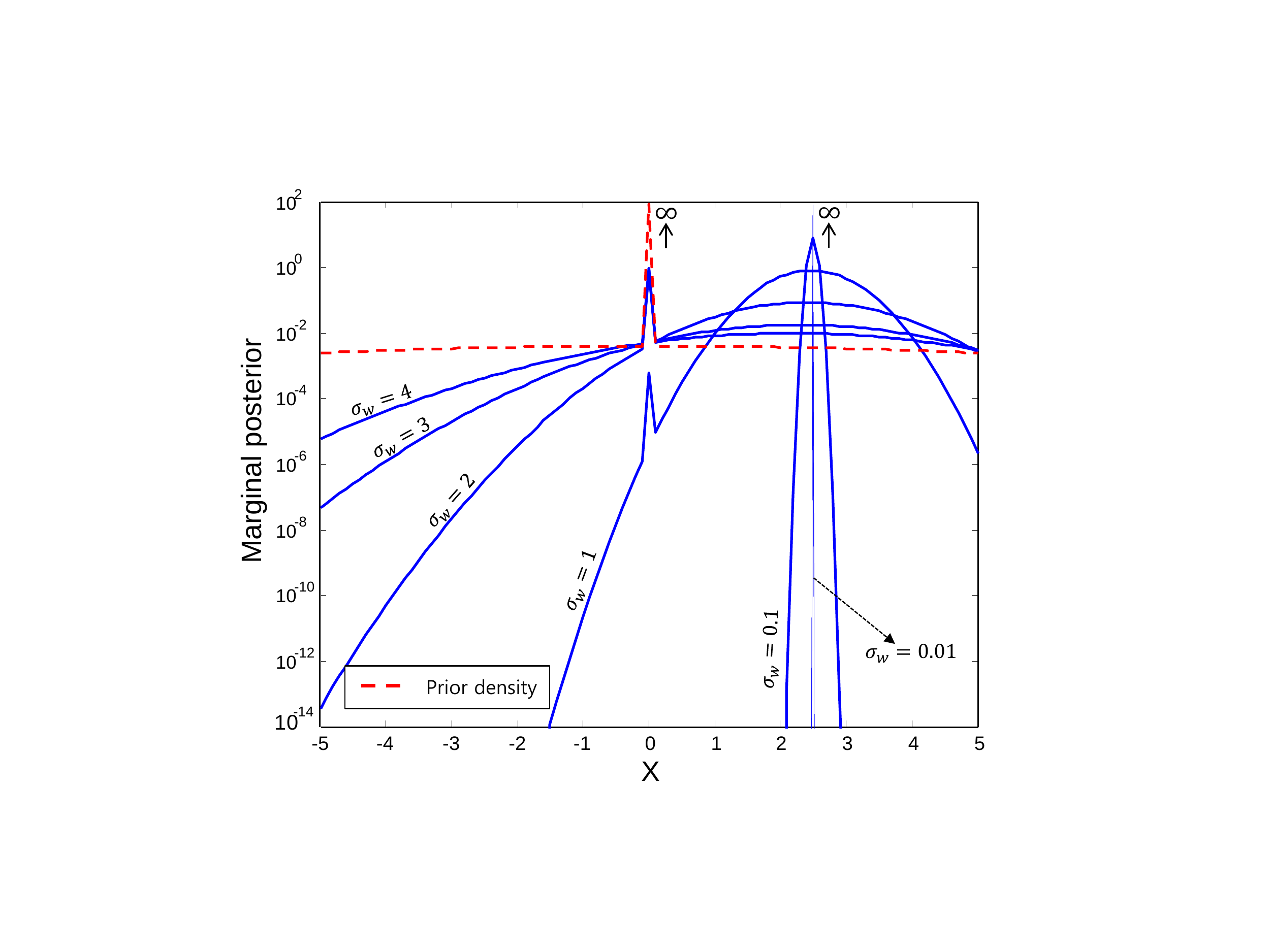}
\caption{Marginal  posterior  corresponding to noise level
$\sigma_W$ when $q=0.05$, $\sigma_X=5$, $x_{0,i}=2.5$, and $L=4$
given $S_i=1$. The figure shows that  the probability mass in the
posterior spreads over the $x$-axis and approaches the prior PDF, as
$\sigma_W$ increases.} \label{fig:fig3-1}
\end{figure*}

\subsection{Phase Transition Analysis of Sparse Support Detection}
We start the phase transition (PT) analysis from studying the
behavior of the marginal posterior $f_{X_{i}}(x|\mathbf{Y},{\bf{\Phi
}})$ corresponding to the noise level $\sigma_W$. We first examine
the two extreme cases: $\sigma_W \rightarrow 0$ and $\sigma_W
\rightarrow \infty$. In the noiseless setup, \emph{i.e.},
$\sigma_W\rightarrow0$, the parameters of
$f_{X_{i}}(x|\mathbf{Y},{\bf{\Phi }})$ converges to $\mathop {\lim
}\limits_{{\sigma _W} \to 0 } {\rho_{i}} = 1, \mathop {\lim
}\limits_{{\sigma _W} \to 0} {\mu _{i}} =x_{0,i}, \mathop {\lim
}\limits_{{\sigma _W} \to 0} {\theta _{i}} = 0$. Therefore, the
marginal posterior converges to a delta function peaked at
$x_{0,i}$, \emph{i.e.},
\begin{eqnarray}\label{eq:eq3-8}
\mathop {\lim }\limits_{{\sigma _W} \to 0
}f_{X_{i}}(x|\mathbf{Y},{\bf{\Phi }}) = \delta_{x_{0,i}}.
\end{eqnarray}
As $\sigma_W$ increases, the probability mass in the posterior
spreads over the $x$-axis, as shown in Fig.\ref{fig:fig3-1}. In the
extreme case, \emph{i.e.}, $\sigma_W\rightarrow\infty$, then the
parameters converges to $\mathop {\lim }\limits_{{\sigma _N} \to
\infty } {\rho_{i}} = q, \mathop {\lim }\limits_{{\sigma _W} \to
\infty} {\mu _{i}} =0, \mathop {\lim }\limits_{{\sigma _W} \to
\infty} {\theta _{i}} = \sigma_X$; hence, the marginal posterior
converges to the prior PDF, \emph{i.e.},
\begin{eqnarray}\label{eq:eq3-9}
\mathop {\lim }\limits_{{\sigma _W} \to \infty
}f_{X_{i}}(x|\mathbf{Y},{\bf{\Phi }}) =f_{X_i}(x).
\end{eqnarray}
Note that the results of \eqref{eq:eq3-8} and \eqref{eq:eq3-9} are
independent of the other parameters ${\sigma _{{X}}},x_{0,i},q$.
These facts indicate that, when the measurements $\mathbf{Y}$ are
clean, the detection of the supportive state of $x_{0,i}$ can be
achieved by either approach given in \eqref{eq:eq2-6} or
\eqref{eq:eq2-9} without uncertainty, whereas in the extremely noisy
case, the measurements $\mathbf{Y}$ are not at all useful for
recovery of $x_{0,i}$, and the both approaches do not work for the
support detection.

We now consider the failure event of the support detection. Let
$\mathcal{F}_i$ denote the failure event  of an index $i \in
\{1,...,N\}$, \emph{i.e.},
\begin{align}\label{eq:eq3-9-2}
\mathcal{F}_i:=\{ \text{Decide}\,
 \mathcal{H}_1,  S_i=0 \} \cup \{ \text{Decide}\, \mathcal{H}_0,
 S_i=1\},
\end{align}
which is the union of two possible failure cases in elementwise
support detection. Let us  consider the first case  $\{
\text{Decide}\, \mathcal{H}_1,  S_i=0 \}$ in \eqref{eq:eq3-9-2}. For
any zero element, \emph{i.e.}, $ S_i=0$, it can be shown that the
additive noise does not affect the success of the elementwise
support detection. Namely, the  state detection of $S_i=0$ is always
right regardless of the noise level. $i)$ When the situation is
noiseless, we know from \eqref{eq:eq3-8} that the marginal posterior
becomes the delta function peaked at the true value $x_{i,0}=0$.
Therefore, it is obvious  that
\begin{align}\label{eq:eq3-10-1}
\arg \mathop {\max }\limits_x \mathop {\lim }\limits_{{\sigma _N}
\to 0} {f_{{X_i}}}(x|{\bf{Y}},{\bf{\Phi }},{S_i} = 0) &= \arg
\mathop {\max }\limits_x {\delta _0} \nonumber\\&= 0,
\end{align}
where the second line holds owing to the nature of the delta
function $\delta_0$. $ii)$ The challenging case is when the noise
level $\sigma_W$ is large. But, from \eqref{eq:eq3-9}, we have
already seen that the marginal posterior converges to the prior
density as $\sigma_W$ increases. Hence, clearly we have
\begin{align}\label{eq:eq3-10-2}
\arg \mathop {\max }\limits_x \mathop {\lim }\limits_{{\sigma _N}
\to \infty } {f_{{X_i}}}(x|{\bf{Y}},{\bf{\Phi }},{S_i} = 0) &= \arg
\mathop {\max }\limits_x {f_{{X_i}}}(x) \nonumber\\&= 0,
\end{align}
where the second line holds by definition of the prior density given
in \eqref{eq:eq2-3}. From $i)$ and $ii)$, the peak of
$f_{X_{i}}(x|\mathbf{Y},{\bf{\Phi }},{S_i} = 0)$ remains at $x=0$
regardless of the noise level, meaning that the state of any zero
element having $S_i=0$, is detected perfectly with no failure,
\emph{i.e.}, $\{ \text{Decide}\,\mathcal{H}_1, S_i=0\}=\emptyset$.
Therefore, the failure event $\mathcal{F}_i$ is confined to the case
$ S_i=1$, \emph{i.e.},
\begin{align}\label{eq:eq3-9-3}
\mathcal{F}_i= \{ \text{Decide}\, \mathcal{H}_0, S_i=1\}.
\end{align}

This result in \eqref{eq:eq3-9-3} reveals an important fact which
the additive noise only disturbs the detection of signal elements on
the support set. Such a result was also discussed in
\cite{tropp},\cite{Donoho}, in terms of the  $l_1$-norm recovery and
OMP. Returning to \eqref{eq:eq3-9-3}, it is worthwhile to note that,
the result is valid for both BHT-BP and CS-BP because both are
derived from the use of the marginal posteriors.


\begin{figure*}[!]
\centering
\includegraphics[width=16cm]{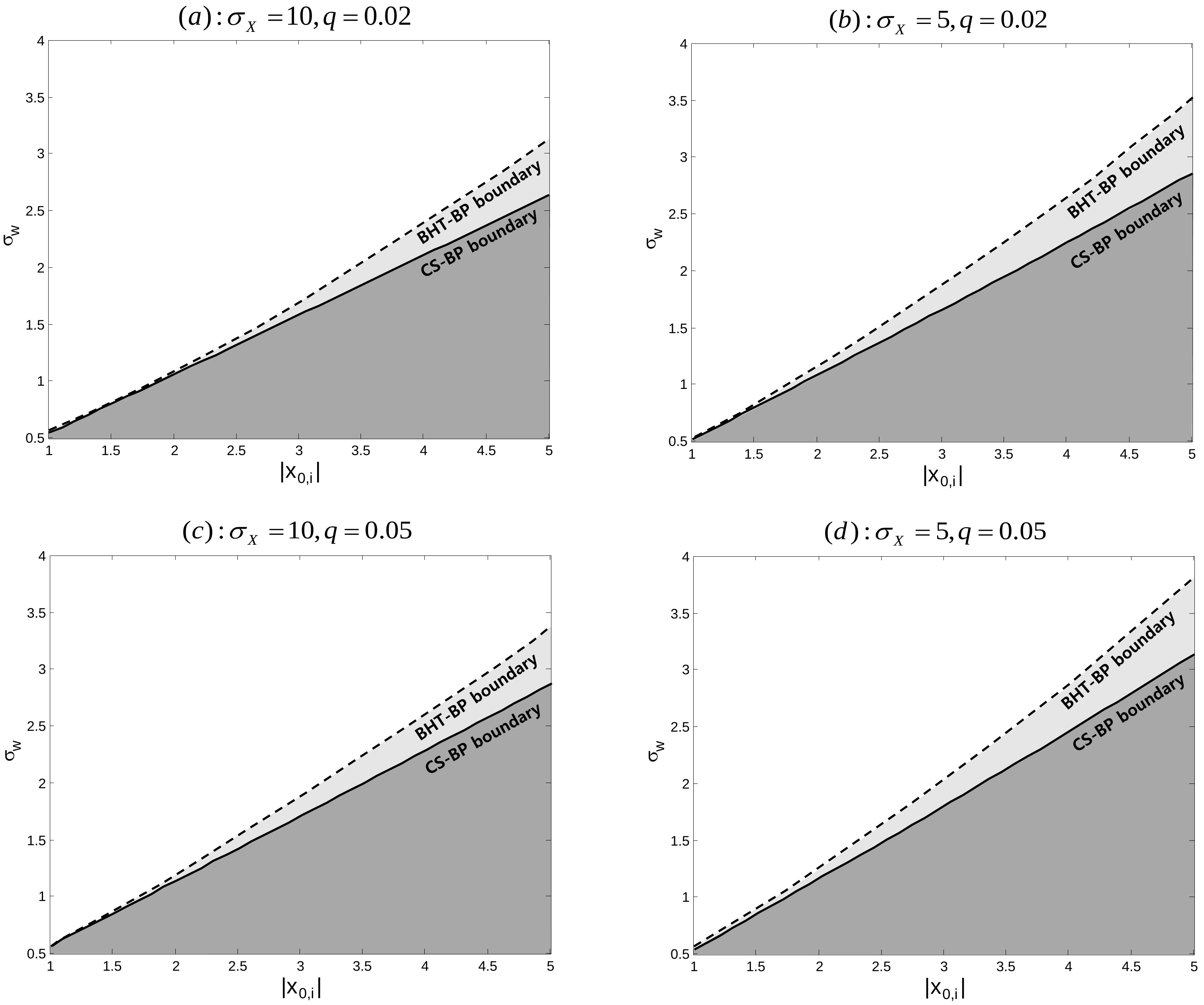}
\caption{PT diagram for elementwise support detection for various
parameter sets of $(\sigma_X,q)$ where the matrix $\mathbf{\Phi}$
with $L=4$ is considered. The dashed curve and solid curve indicate
BHT-BP and CS-BP, respectively. In these figures, the region above
the curves corresponds to the SSD-failure and the region below
corresponds to the SSD-success.} \label{fig:fig4-7}
\end{figure*}

In order to draw the PT boundary of BHT-BP and CS-BP, we need to
find the condition which causes the failure event $\mathcal{F}_i$
\eqref{eq:eq3-9-3} of each detector, with respect to  the noise
level $\sigma_W$ and the signal magnitude $|x_{i,0}|$ on the signal
support. For each element on the support, the failure event is
equivalent to the case when the detection function is nonpositive,
\emph{i.e.},
\begin{align}\label{eq:eq3-12}
h_{\text{CS-BP},i}(\sigma_W,x_{0,i})\leq 0,
\end{align}
for CS-BP from \eqref{eq:eq2-9}, and
\begin{align}\label{eq:eq3-11}
h_{\text{BHT-BP},i}(\sigma_W,x_{0,i})\leq 0,
\end{align}
for BHT-BP from  \eqref{eq:eq2-6},  respectively. Note that
$h_{\text{BHT-BP},i}$ and $h_{\text{CS-BP},i}$ are  functions of the
noise level $\sigma_W$ and square of the signal element $x_{0,i}^2$.
Therefore, we  can ignore the sign of $x_{i,0}$ and  handle the
signal magnitude $|x_{0,i}|$ in this analysis. The equality
condition of \eqref{eq:eq3-11}, and that of \eqref{eq:eq3-12} divide
the plane of $\sigma_W$ and $|x_{0,i}|$ into two distinct regions:
an `SSD-failure' and an `SSD-success', as depicted in
Fig.\ref{fig:fig4-7}.  In the figure, the region above the boundary
corresponds to $h(\sigma_W,x_{0,i}) \leq 0$, which describes the
failure, whereas the region below corresponds to
$h(\sigma_W,x_{0,i}) > 0$, which describes the success. Such a
figure is what we mean by the term PT analysis. Hence, the boundary
derived from the equality condition determines the region of success
and failure over the plain of $\sigma_W$ and $|x_{0,i}|$. The
broadness of the success region indicates the recovery ability of
the corresponding detector. Namely, the wider the success region is,
the more capable a detector is.

The BHT-BP boundary  provides a wider success region than the CS-BP
boundary does. Examples are shown  in Fig.\ref{fig:fig4-7}-(a) for
the parameter set $\sigma_{X}=10, q=0.02$, Fig.\ref{fig:fig4-7}-(b)
for $\sigma_{X}=5, q=0.02$, Fig.\ref{fig:fig4-7}-(c) for
$\sigma_{X}=10, q=0.05$, and Fig.\ref{fig:fig4-7}-(d) for
$\sigma_{X}=5, q=0.05$ where we consider the matrix $\mathbf{\Phi}$
with $L=4$. From these examples,  it is evident that BHT-BP is
superior to CS-BP. In these figures,  we also note that BHT-BP
generally performs better when  the signal magnitude $|x_{0,i}|$ is
larger.

\begin{figure*}[!t]
\centering
\includegraphics[width=16cm]{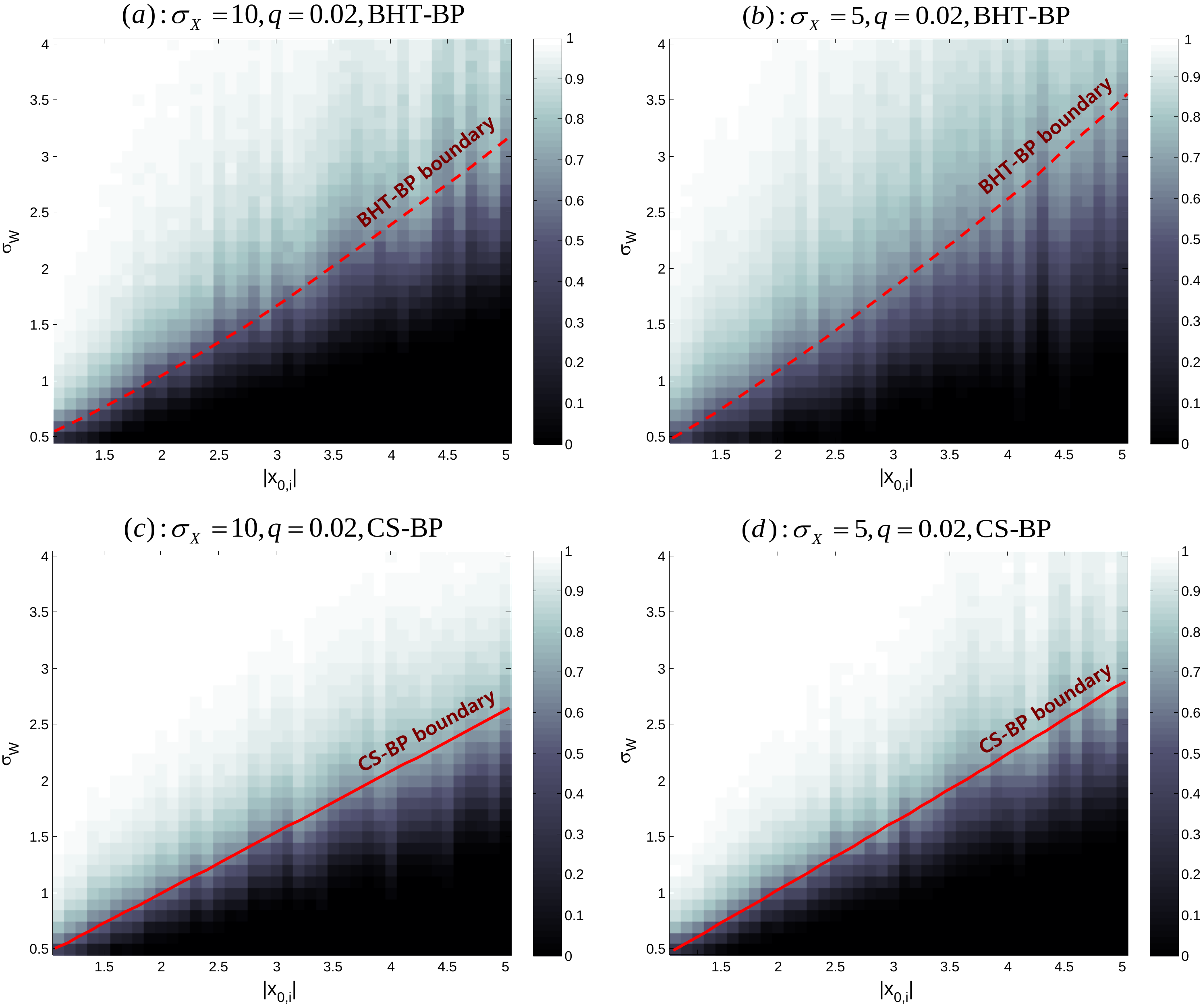}
\caption{Experimental probability of support detection failure for
BHT-BP and CS-BP when  $N=1024, q=0.02$, and $L=4$.}
\label{fig:fig4-10}
\end{figure*}

The result shown in Fig.\ref{fig:fig4-7} can be interpreted as
follows. In the case of CS-BP, additive noise produces the event
$\mathcal{F}_i$ when the zero-spike becomes the peak of the marginal
posterior by exceeding the peak of the slab-PDF, as described in
\eqref{eq:eq2-9}. For example, CS-BP misdetects the supportive state
with the posteriors of $\sigma_W \geq 2$ given in
Fig.\ref{fig:fig3-1}. In the BHT-BP case, however, the detector
determines the supportive state by considering the noise spreading
effect of the posterior, which corresponds to the inner products of
the marginal posterior and the function consisting of the prior
knowledge, \emph{i.e.}, ${\int { \frac{{f_{X_i}( x |S ) }}{{f_{X_i}(
x ) }} f_{X_i}(x | {\bf{Y}},{\bf{\Phi }} ) }dx } $, as described in
\eqref{eq:eq2-6}. Hence, the detection of BHT-BP is performed by
incorporating the posterior function $ f_{X_i}(x |
{\bf{Y}},{\bf{\Phi }} )$ over the entire $x$-axis, in contrast to
that of CS-BP which only considers the function at a given point.
Therefore, BHT-BP  detects the signal support against additive noise
more robustly than CS-BP does.

\begin{figure*}[!t]
\centering
\includegraphics[width=16cm]{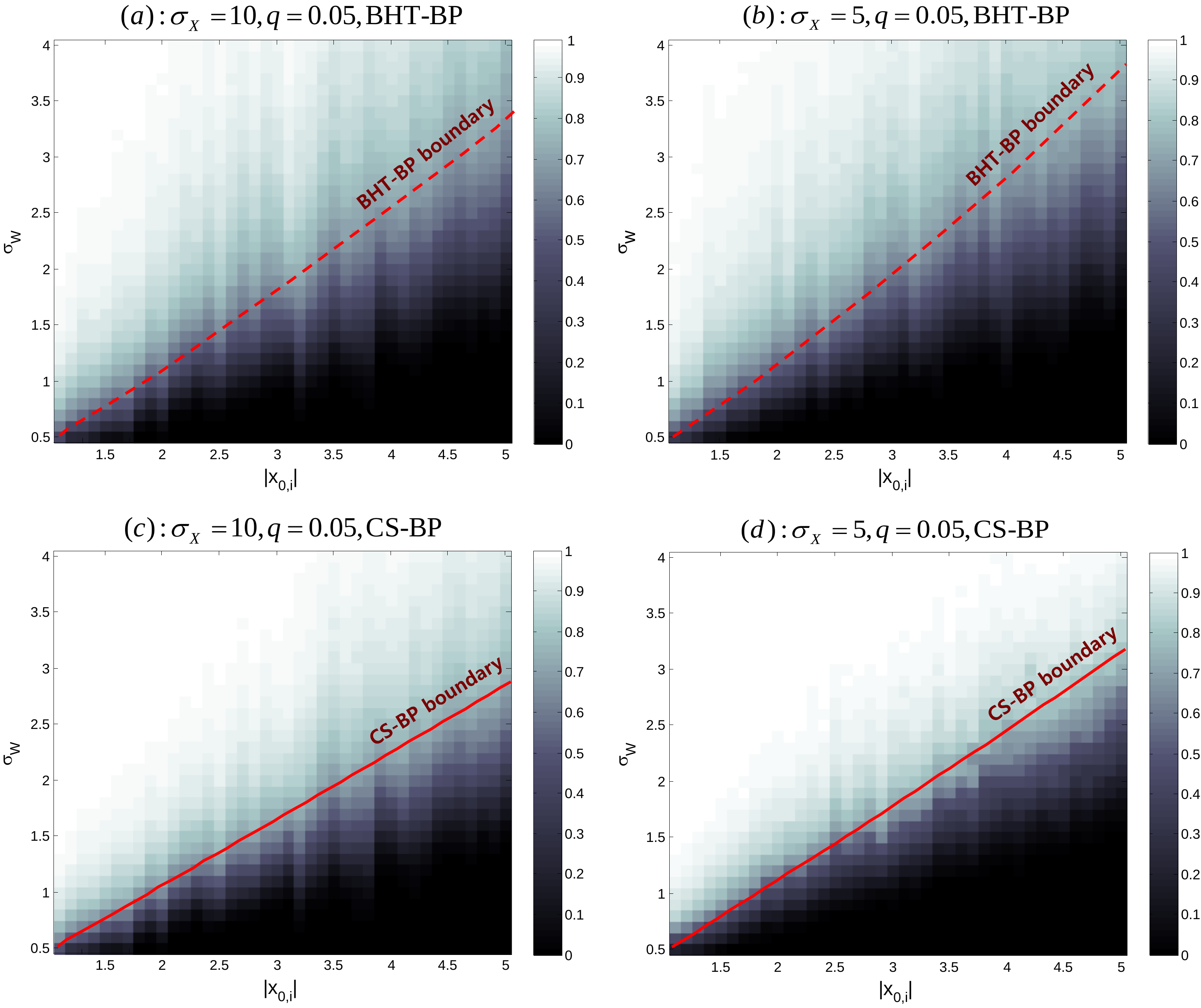}
\caption{Experimental probability of support detection failure for
BHT-BP and CS-BP when  $N=1024, q=0.05$, and $L=4$.}
\label{fig:fig4-11}
\end{figure*}

\section{Experimental Results}
We support our analytical results with an experimental validation.
We measured the probability of the SSD-failure, defined as
\begin{align}\label{eq:eq4-1}
\Pr \{\mathcal{F}_i |S_i=1\},
\end{align}
for both BHT-BP and CS-BP over the plain of the noise level
$\sigma_W$ and  signal magnitude $|x_{0,i}|$. For the evaluation of
each experimental point, we used the Monte Carlo method with 100
trials where each trial is generated under the parameters of
$N=1024, L=4$ and the undersampling ratio $N/M=0.5$. To obtain  the
marginal posteriors of the signal, we used the iterative BP
algorithm introduced in \cite{CS-BP},\cite{SSP2012},\cite{BHT-BP}.

The experimental failure probability of BHT-BP and CS-BP is shown in
Fig.\ref{fig:fig4-10} and Fig.\ref{fig:fig4-11} for various
parameter sets  of $(\sigma_X, q)$, with the corresponding
analytical boundary given in Fig.\ref{fig:fig4-7}. In these figures,
the brightness of each experimental point represents the value of
the failure probability. For example, when the color is bright, the
failure probability is close to one. These experimental results show
that the figures of CS-BP include a wider white region than that of
BHT-BP.  We also note in these figures that the transition indeed
occurs near the analytical boundary. Thus, we can say that these
experimental results are in good agreement with the analytical
results given in Section III-B.

\section{Conclusion}
The main aim of this investigation is to answer if a detection
oriented algorithm BHT-BP \cite{SSP2012},\cite{BHT-BP} provides any
better results than an estimation based algorithm CS-BP \cite{CS-BP}
for the recovery of sparse support in a noisy underdetermined system
of equations. A focus is to come up with a measure with which
superiority of one to the other can be made precisely. To this end,
we first obtained an expression of the marginal posterior as a
function of the noise level and signal magnitude. Using the
posterior expression, we have shown that the support detection
errors occur only for signal elements on the support. We used such a
fact to find the PT boundary which divides the plane of the noise
level and signal magnitude into two distinct regions: an
``SSD-failure" and an ``SSD-succes." Specifically, the diagram
provides information on the required signal magnitude for the
recovery under a certain noise level or the allowable noise level
for the recovery given a fixed signal magnitude. The PT analysis
shows in a clear transition diagram how much BHT-BP is better than
CS-BP, in recovering the sparse support.

%
%
%
%



\begin{thebibliography}{10}
\bibitem{tropp} J. A. Tropp, ``Just relax: convex programming methods for
identifying sparse signals in noise," \emph{IEEE Trans. Inform.
Theory}, vol. 52, no. 3, pp. 1030-1051, 2006.

\bibitem{Lasso} R. Tibshirani, ``Regression shrinkage and selection via the
lasso," \emph{J. Roy. Statisti. Soc., Ser. B}, vol. 58, no. 1, pp.
267-288, 1996.

\bibitem{CS-BP} D. Baron, S. Sarvotham, and R. Baraniuk, ``Bayesian compressive sensing via belief propagation,"
 \emph{IEEE Trans. Signal Process.}, vol. 58, no. 1, pp. 269-280, Jan.
2010.






\bibitem{Wainwright2}  M. J. Wainwright, ``Information-theoretic limits on sparsity
recovery in the high-dimensional and noisy setting," \emph{IEEE
Trans. Inform. Theory}, vol. 55, no. 12, pp. 5728-5741, Dec. 2009.

\bibitem{Akcakaya} M. Akcakaya and V. Tarokh, ''Shannon-theoretic limit on noisy compressive sampling," \emph{IEEE Trans. Inform. Theory}, vol. 56, no. 1, pp.
492-504, Jan. 2010.

\bibitem{Fletcher} A. Fletcher, S. Rangan, and V. Goyal, ''Necessary and sufficient conditions for sparsity pattern recovery," \emph{IEEE Trans. Inform.
Theory}, vol. 55, no. 12, pp. 5758-5772, Dec. 2009.



\bibitem{SSP2012} J. Kang, H.-N. Lee, and K. Kim, ``Bayesian hypothesis test for sparse support recovery using belief propagation,"
\emph{Proc. in IEEE Statistical Signal Processing Workshop (SSP)},
pp. 45-48, Aug. 2012.

\bibitem{BHT-BP} J. Kang, H.-N. Lee, and K. Kim, ``Detection-Directed sparse estimation using Bayesian hypothesis test and belief propagation,"
2012 [Online]. Available: arXiv:1211.1250 [cs.IT].

\bibitem{PT1} D. L. Donoho and J. Tanner, ``Neighborliness of randomly
projected simplices in high dimensions," \emph{Proc. Nat. Acad.
Sci.} USA, vol. 102, no. 27, pp. 9452-9457, 2005.


\bibitem{PT2} D. L. Donoho and J. Tanner, ``Precise undersampling
theorems," \emph{Proceeding of the IEEE}, vol. 98, issue 6, pp.
913-924, June, 2010.




\bibitem{Guo05} D. Guo and S. Verdu, ``Randomly spread CDMA:
Asymptotics via statistical physics," \emph{IEEE Trans. Inform.
Theory}, vol. 51, no. 6, pp. 1983-2010, Jun. 2005.

\bibitem{Guo07} D. Guo and C.-C. Wang, ``Random sparse linear systems observed via arbitrary
channels: a decoupling principle,"\emph{Proc. in IEEE Int. Symp.
Inform. Theory (ISIT)}, pp. 946-950, Nice, France, June. 2007.

\bibitem{Guo06} D. Guo and C.-C. Wang, ``Asymptotic mean-square optimality of
belief propagation for sparse linear systems," \emph{ Proc. in IEEE
Inform. Theory Workshop (ITW)}, pp. 194-198, Chengdu, China, Oct.
2006.


\bibitem{Donoho} D. L. Donoho, M. Elad, and V. Temlyakov, ``Stable
recovery of sparse overcomplete representations in the presence of
noise," \emph{IEEE Trans. Inform. Theory}, vol. 52, no. 1, pp. 6-18,
Jan. 2006.





%
%
%
%
%

%







\end{thebibliography}
\end{document}